\documentclass[prl,noshowkeys,twocolumn,superscriptaddress]{revtex4}
\usepackage{graphicx}
\usepackage{amsfonts}
\usepackage{amsmath}

\usepackage[latin1]{inputenc}
\usepackage{pstricks,pst-node,pst-text,pst-3d,pslatex}
\usepackage[T1]{fontenc}

\usepackage{graphicx}
\usepackage{amsfonts}
\usepackage{amsmath}

\newcommand{\be}{\begin{eqnarray}}
\newcommand{\ee}{\end{eqnarray}}

\newcommand{\comment}[1]{}

\begin{document}
\author{M.~Sega}
\affiliation{C.N.R./I.N.F.M. and Dipartimento di Fisica, Universit\'a degli Studi di Trento, Via Sommarive 14, Povo (Trento), I-38050 Italy.} 
\author{P.~Faccioli}
\affiliation{Dipartimento di Fisica  Universit\'a degli Studi di Trento e I.N.F.N, Via Sommarive 14, Povo (Trento), I-38050 Italy.} 
\affiliation{European Centre for Theoretical Studies in Nuclear Physics and Related Areas (E.C.T.$^*$), Strada delle Tabarelle 284, Villazzano (Trento), I-38050 Trento.}
\author{F.~Pederiva}
\affiliation{Dipartimento di Fisica and C.N.R./I.N.F.M.-DEMOCRITOS National Simulation Center,  Universit\'a degli Studi di Trento, Via Sommarive 14, Povo (Trento), I-38050 Italy.}
\author{G.~Garberoglio}
\affiliation{C.N.R./I.N.F.M. and Dipartimento di Fisica, Universit\'a degli Studi di Trento, Via Sommarive 14, Povo (Trento), I-38050 Italy.} 
\author{ H.~Orland}
\affiliation{Service de Physique Th\'eorique,
Centre d'Etudes de Saclay, F-91191, Gif-sur-Yvette Cedex, France.}
\title{Quantitative Protein Dynamics from Dominant Folding Pathways}

\begin{abstract}
We develop a theoretical approach to the protein folding problem based on out-of-equilibrium stochastic dynamics. Within this framework, the computational difficulties related to the existence of large time scale gaps in the protein folding problem are removed and simulating the entire reaction  in atomistic details  using existing computers becomes feasible.  
In addition, this formalism provides a natural framework to investigate the relationships between  thermodynamical and kinetic aspects of the folding. For example, it is possible to show that, in order to have a large probability to remain unchanged under Langevin diffusion, the native state has to be characterized by a small conformational entropy. We discuss how to determine  the most probable folding pathway, to identify configurations representative of the transition state and to compute the most probable transition time. 
 We perform an illustrative application of these ideas, studying the conformational evolution of alanine di-peptide, within an all-atom model based on the empiric GROMOS96 force field. 
\end{abstract}

\maketitle

A critical part of the protein-folding problem is to understand its 
kinetics and the underlying physical processes. To this
aim, several different theoretical methods have been recently developed,
spanning from  analytical approaches\cite{TheoAdv1,TheoAdv2,TheoAdv3} to detailed
computer simulations\cite{SimAdv1,SimAdv2,SimAdv3}. A major problem in simulating
the folding process using standard molecular dynamics (MD) is the huge
gap between the time scale of ``elementary
moves'', of the order of 10-100 ps,  and that of the entire folding process,
which ranges from a few microseconds for fast-folders\cite{FastFolders}, up to several 
seconds or even minutes for more complex proteins. This peculiarity
of the folding process makes the brute-force molecular dynamics 
approach too demanding, and a substantial part of the efforts in the
field of protein folding simulation aims at
bridging this gap.

In a recent paper~\cite{DFP} we have presented a novel theoretical
framework for investigating the folding dynamics, named hereafter 
Dominant Folding Pathways (DFP), which is based on a reformulation
in terms of path integrals of the dynamics described by the Langevin
equation. The DFP analysis allows to compute rigorously 
(i.e. without any assumptions other than the validity of the underlying
Langevin equation) the most probable conformational pathway connecting
an arbitrary initial conformation to an arbitrary final conformation. 
The major advantage of
the method is the possibility of bypassing the computational
difficulties associated with the existence of different time scales
in the problem, while retaining the ability to recover information
on the time evolution of the system.  
As we shall see, the resulting computational simplification is dramatic 
and makes it feasible to study the formation pattern of conformational structures along 
the entire folding process using realistic all-atom force fields, on available computers.

In this Letter we further develop our
formalism and we present the first DFP simulation performed in full atomistic detail.
We show how the DFP analysis gives access to
important information about the dynamics of the folding process,
such as the characterization and determination of the transition state,
and the most probable transition time.
In addition, we show that in this  formalism the  native state is characterized by a single effective parameter and this leads to an interesting relationship between kinetic and thermodynamical quantities.

Let us begin our discussion by briefly reviewing the key concepts
of the DFP method,  here presented for a simple one-dimensional system,
without loss of generality.

The DFP method can be applied to any system described by the over-damped Langevin
equation 
\begin{equation} \label{Langevin}\frac{\partial x}{\partial t} = -\frac{D}{k_B
T} \frac{\partial U}{\partial x} + \eta(t), 
\end{equation} 
where $U$ is the potential energy of the system, $\eta(t)$ is a Gaussian
random force with zero average and correlation given by
$\langle\eta(t)\eta(t')\rangle= 2D\delta(t-t')$. Note that in the original Langevin equation there is a mass term, $m \ddot{x}$. However, as shown in \cite{Pitard}, for proteins, this term can be neglected beyond time scales of the order of $10^{-13}$ s.

The probability of finding the system in a conformation $x_f$ at time $t_f$ starting from
a conformation $x_i$ at $t_i$, is a solution of the well-known Fokker-Planck
Equation, and can be expressed as a path-integral:
\be
\label{path}
P(x_f,t_f|x_i,t_i)=e^{-\frac{U(x_f)-U(x_i)}{2 k_B T}}
\int_{x_i}^{x_f} \mathcal{D}x(\tau)\, 
e^{-{S_{eff}[x]}},
\ee
where 
\be
S_{eff}[x]=\int_{t_i}^{t_f} d\,\tau~ 
\left(\frac{\dot{x}^2(\tau)}{4 D}+ V_{eff}[x(\tau)]\right),
\ee
is called the effective action and 
 \be
\label{veff}
V_{eff}(x)=\frac{D}{4(k_B T)^2}\,\left[\left(\frac{\partial
U(x)}{\partial x}\right)^2
-2k_B T\frac{\partial^2~U(x)}{\partial x^2}\right].
\ee
is called the effective potential.
This quantity measures the tendency of a configuration to evolve under Langevin diffusion. In fact, the probability for the system to remain in the same configuration $x$ under an infinitesimal time interval is given by
\be
P(x,dt|x,0)= e^{-V_{eff}(x)dt}
\label{Pxx}
\ee 
Hence, points of large effective potential are highly unstable under Langevin diffusion.

There is an obvious sum rule: $\int dx_f P(x_f,t_f|x_i,t_i) =1 $,
which can be written as
\be
1= e^{\frac{U(x_i)}{2 k_B T}} \int d x_f~\int_{x_i}^{x_f} \mathcal{D}x(\tau)\, 
e^{-\frac{U(x_f)}{2 k_B T} -{S_{eff}[x]}}
\label{sumrule}
\ee
>From a saddle-point analysis of this sum-rule, it follows that 
given the initial condition $(x_i,t_i)$,
the most probable
paths contributing to (\ref{sumrule}) satisfy the 
Euler-Lagrange equations derived from the effective Lagrangian
$L_{eff}=\dot{x}^2/4 D + V_{eff}(x)$ with proper boundary conditions
\begin{eqnarray}
\frac{d^2 x}{dt^2} &=& 2D \frac{\partial V_{eff}}{\partial x} \\  
\dot{x}_f &=&- \frac{D}{k_B T} U'(x_f) \label{end}\\
x(t_i) &=& x_i
\end{eqnarray}
The saddle-point equation (\ref{end}), which comes from the variation of the exponent in (\ref{sumrule}) with respect to the final point $x_f$,
tells us which final points dominantly contribute  to the sum rule, that is which conformations are most likely to be visited
 at the final time, $t_f$.

The numerical advantage in determining the DFP comes from the fact 
that $L_{eff}$ is a Lagrangian
describing  an energy conserving dynamics, and therefore it is possible to use the Hamilton-Jacobi
(HJ) description. With this change of framework, the total computational
cost of the simulation now depends on the length of the path, rather than
on the folding time.  In the HJ framework, the most probable pathway
is obtained by minimizing ---~not just extremizing~--- the functional
\be S_{HJ}([x];x_i,x_f)=~\int_{x_i}^{x_f} dl
\sqrt{1/D\left(E_{eff}+V_{eff}[x(l)]\right)},
\ee where $dl=\sqrt{(d x)^2}$
is an infinitesimal conformational change along the path and the effective energy is given by
\be
E_{eff}= \frac{\dot{x^2}}{4 D} -V_{eff}(x).
\ee
Since the effective energy is conserved along the DFP, using equation (\ref{end})  we have
\be
\label{energy}
E_{eff} =\frac{D}{2 k_B T} U''(x_f). 
\ee
Note that, since the diffusion coefficient drops exponentially at low temperatures,  $D=D(T) \sim D_0 \exp[-E_a/k_B T]$, the effective energy vanishes in this limit.
In the long time limit, $t_f  \to \infty$, we know that $P(x_f,t_f|x_i,t_i) $ converges to the Boltzmann distribution
\be
P(x_f,t_f|x_i,t_i) \to_{t_f \to \infty} \frac{e^{-U(x_f)/k_B T}}{Z}
\ee
where $Z$ is the partition function of the system.
If the reaction takes place at a temperature below the folding temperature,
at large times  the system will sample configurations $x_f$ close to the global minimum-potential-energy configuration  $x_n$, 
for which $U'(x_n)=0$.

We shall define the {\it native state} as the region of configuration space which is thermally accessible from the minimum-energy conformation $x_n$, i.e. for which energy differences with respect to $U(x_n)$ are of the order of $k_B T$.
We can assume that, for all configurations $x_f$ in the native state, the potential energy can be described in the harmonic approximation:
\be
U(x_f) \approx U(x_n) + \frac{1}{2} U''(x_n) (x_f-x_n)^2
\ee
This equation together with equation (\ref{energy}) implies
\be
\label{power}
E_{eff}= \frac{D(T)}{2 k_B T} U''(x_n)= - V_{eff}(x_n)
\ee
This equation is quite powerful, since it shows that the effective energy does not depend on the specific conformation $x_f$ in the native state, and is totally determined by the temperature of the heat-bath and by the curvature of the potential energy at the minimum-energy point $x_n$. Stated differently, the native state belong to a surface in configuration space of constant effective energy $E_{eff}$. 

This parameter appears in the macroscopic quantities characterizing the thermodynamics of the native state. As an example, 
%let us consider the {\it conformational entropy} of the native state, which  can be defined as the number of micro-states in the native state, i.e. 
let us discuss the  conformational entropy $S_{conf}$, which  measures the number of micro-states in the native state, i.e. the contribution to the partition function of all the configurations for which $U(x)-U(x_n)\lesssim k_B T$, where $x_n$ is the minimum-energy configuration: 
\be
Z_N(T) &=& \int d x  e^{-\frac{U(x)}{k_B T}}~\theta\left(k_B T - (U(x)-U(x_n))\right) \\
S_{conf}(T) &\equiv&- \frac{\partial}{\partial T}  k_B T \ln Z_N(T)
\ee
Expanding the potential energy quadratically around the minimum-energy configuration $x_n$ we have 
\be
Z_N &\simeq& e^{-\frac{U(x_n)}{k_B T}}~\int^{x_n+\sqrt{2 k_B T/ U''(x_n)}}_{x_n-\sqrt{2 k_B T/ U''(x_n)}} d x~e^{- \frac{U''(x_n)}{2 k_B T} (x-x_n)^2},
\ee
which gives
\be
S_{conf}(T)&\simeq& k_B  \ln \left[\sqrt{\frac{2 \pi k_B T}{U''(x_n)}}~ \textrm{Erf}(1)\right]\\
&\simeq& k_B \ln \left[\sqrt{\frac{D \pi }{E_{eff}}}\right]+ \textrm{const}.
\label{Sconf}
\ee
This equation  expresses the intuitive fact that the conformational entropy of the native state is small if  the minimum of the potential energy is very narrow. From Eq.s~(\ref{Sconf}) and (\ref{power}), we can conclude that the effective potential $V_{eff}(x)$ provides a measure of the conformational entropy of any (meta)-stable state.

The effective potential also governs the kinetics of the folding.
As a result, in this formalism, it is possible to investigate the relationship between thermodynamical and kinetic aspects of the protein folding reaction. For example, we now show that the stability of the native state is related to its conformational entropy. To this end, let us consider the probability for  the native state to remain unchanged during an elementary time interval $dt$, i.e. the probability that all points in the native state evolve into points which are still in the native state:
\be
&&P(\textrm{native}, dt| \textrm{native},0) \equiv   \int_{U(x)-U(x_n)< k_B T} d x
\nonumber\\
&&~\int_{U(y)-U(x_n)< k_B T} d y ~P(y,dt|x,0) 
\ee
This quantity, which generalizes the persistence probability (\ref{Pxx}),  can be evaluated using Eq.~(\ref{path}) and expanding the exponent in the Gaussian approximation. The result, which is quite involved and will not be presented here, shows that the persistence probability of the native state increases for large local curvatures of the potential energy near the native state, i.e. for $V_{eff}(x)\to -\infty$.
Hence, $V_{eff}(x)$ controls both the stability and the conformational entropy of the native state.
This  implies that, in order to have a large probability to remain unchanged under Langevin diffusion, the native state has to be characterized by a small conformational entropy. 

In the case of a protein with Hamiltonian $H(\vec r_1,\dots,\vec r_N)$, denoting the coordinates of the minimum-energy conformation
 by $\{ {\vec r_I^{(n)}}\}$, and assuming again that in the final stage of the folding, the protein samples the native state, we write the quadratic expansion around the minimum of $H$ as
\begin{eqnarray}
H(\vec r_1,\dots,\vec r_N) =  H(\vec r_1^{(n)},\dots,\vec r_N^{(n)}) \\
+ \sum_{i j,\mu \nu}~\frac{1}{2} (r_i^{\mu}- r_i^{\mu (n)}) \frac {\partial^2 H}{\partial r_i^{\mu (n)} \partial r_j^{\nu (n)}} (r_j^{\nu}- r_j^{\nu (n)})
\end{eqnarray}
This equation implies the equivalent of equation (\ref{power})
\begin{eqnarray}
E_{eff}&=& 
%\frac{D}{2 k_B T}  \sum_i {\vec \nabla^2_i}H \\ &=& 
\frac{D}{2 k_B T} \sum_i {\vec \nabla^2_i}H  (\vec r_1^{(n)},\dots,\vec r_N^{(n)}) \\
&=& \frac{D}{2 k_B T} {\rm Tr} ~\mathcal{H}^{(n)} 
\end{eqnarray}
where $\mathcal{H}^{(n)}$ is the Hessian matrix around the minimum-energy conformation.
Obviously, such a quantity can be obtained either from a normal mode analysis around the native state, or
equivalently  by evaluating the average of the velocities
$v_i^2\equiv (d{\vec r_i}/dt)^2$
from several short MD simulations around the native state.

To summarize the strategy to find the most probable reaction paths, we may proceed as follows:
(i)~Prepare several initial denatured conformations by running  short MD simulations at high temperature. (ii)
~Prepare a representative set of the native state by making short time MD simulations from the minimum-energy configuration. These short time MD simulations also allow to compute the trace of the Hessian matrix, and thus the effective energy $E_{eff}$.
(iii) Solve the Hamilton-Jacobi equations from the denatured conformations to the native conformations, using the energy $E_{eff}$ computed above.

%In our previous work\cite{DFP}, we employed a very simple representation
%GO-model) of a protein fragment to test the simulation protocol. 
In order to make quantitative predictions on the folding process,  we need to show that this framework 
can be successfully applied to all-atom models, using available computers.
As a first application of this type,  we study the kinetics of alanine dipeptide, which is usually 
the benchmark system for the investigation of new simulation methods in this field.\cite{Ala1,Ala2,Ala3}.
The force-field employed is GROMOS96~\cite{Gromos96}, while the electrostatics
effects mediated by the solvent are accounted for by imposing a dielectric
permittivity $\epsilon_r=80$,  leaving more sophisticated
implicit descriptions of the solvent to forthcoming phenomenological
applications. 
%In this approximation, the free energy landscape
%can be easily computed by direct integration.  

\begin{figure}
\vspace{0.2 cm}
\includegraphics[width=\columnwidth]{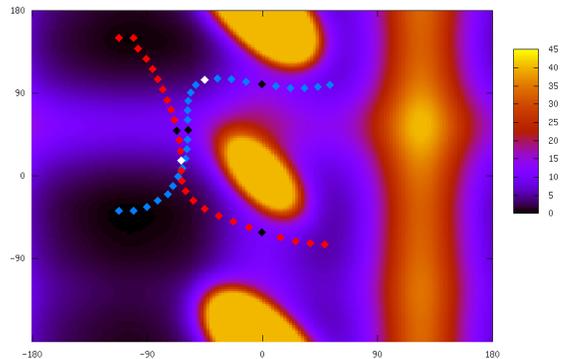}
\caption{
Dominant Folding Paths for the C7$_{ax}\to$C7$_{eq}$ (red squares) and
$\alpha_L\to\alpha_R$ (blue squares) transitions. In the background,
the free energy profile for the $\psi$ and $\phi$ dihedrals is shown (in
units of kJ/mol). Black squares identify the minimum residence time
conformations, and the white squares the transition states defined by comittement analysis.}
\label{figres1}
\end{figure}

In Fig. (\ref{figres1}) we present the results of the DFP
analysis relative to two specific transitions (C7$_{ax}\to$C7$_{eq}$ and
$\alpha_L\to\alpha_R$), compared with the Free Energy landscape computed by direct integration.
 The values of the two $\psi$ and $\phi$ dihedrals
along the paths obtained by minimising the effective action are plotted
on top of the relative free energy map.   
These simulations were performed at temperature $T=300K$, and  assuming a
diffusion coefficient $D=0.02$\AA$^2$ps$^{-1}$ for all atoms.  To determine
the DFP we have performed 500 cycles of simulated annealing of the
discretized Hamilton-Jacobi functional~\cite{DFP}, followed by a refinement stage where Conjugate
Gradients were used. The effective energy $E_{eff}$ was estimated  running  a few ps of MD simulation
 starting from the minimum-energy conformation.

We will now show how the DFP analysis can provide valuable information
about the dynamics of the transition, and about the determination
of the transition state along the path.
While other methods with similar purposes, e.g.\cite{SimAdv3,Ala3},
only provide a meta-dynamics, the DFP analysis yields
information on the real-time evolution of the system. Even though time is
no longer an independent variable of the calculation in the HJ formulation,
the total time required to perform the transition from a conformation $x_i$ to a conformation $x_f$ 
can be computed as
\be
\label{time} t_f-t_i=\int_{x_i}^{x_f}dl
\frac{1}{\sqrt{4 D \left(E_{eff}+V_{eff}[x(l)]\right)}}
\ee 
and the
time spent in the neighborhood of each intermediate conformation
(residence time along the path) is easily derived from the differential
form of Eq.~\ref{time}.
The computed times for the C7ax$\to$C7eq and $\alpha_L\to\alpha_R$
transitions are  12.0 and 11.4~ps, respectively.  Notice, that
this is the most probable transition time, and not the mean first
passage, or Kramers time~\cite{carolifunct}.

%For example, given a transition visiting two configurations A and B,
%connected by the transition state T, the time appearing in \ref{time}
%is relative to the transition $ATB$, which passes through T only once. 
%On the other hand, the Kramers time includes also the contribution from paths such as ATATB, ATATATB,\ldots,
%each one weighted with the corresponding probability~\cite{carolifunct}.

An analysis of the residence time along the path shows that in each of the two DFP's,
there are two points where the conformation of alanine dipeptide
has shortest residence time. These points, indicated in Fig.~(\ref{figres1}) with
black symbols, are located in the proximity of the saddle-points of the free
energy landscape, as one would expect. 

On the other hand, within the present formalism it is also possible to rigorously define the {\it transition state} 
along the path in terms of commitment analysis~\cite{bunsen,pandeTS}.
Following Eq. (2), once  the DFP has been determined,
the conformation $x_{ts}$ is easily obtained  
by requiring  that the probability in the saddle point approximation
to diffuse back  to the initial configuration $x_i$,  $P(x_i|x_{ts})$
 equates that  of evolving toward the final native configuration $x_f$, $P(x_f|x_{ts})$.
In the saddle-point approximation, this condition leads to  the simple equation:
\be
\frac{U(x_f)-U(x_i)} {2 k_B T } = S_{HJ}([x];x_{ts},x_i)-S_{HJ}([x];x_{ts},x_f).
\ee
We want to point out that this definition of $x_{ts}$ neither relies
on the use of any specific reaction coordinate, nor on the {\it a priori} knowledge of
the free energy landscape, but is purely based on the properties of
the diffusive dynamics followed by the system. Transition states computed
using this prescription are shown in Fig.\ref{figres1} as white
points. These results provide a clean example of the fact that the 
definition of transition-state in terms of commitment analysis can be used to locate the configuration of highest free-energy barrier only in the case of two-state transitions.

In conclusion, in the present work we have developed a new theoretical description
of the  protein folding reaction, based on  Langevin dynamics.
This approach allows for a huge reduction of the computational cost
needed for obtaining information on the full reaction pathway.
Within this framework,  all-atom simulations for a dipeptide
can be performed in just a few minutes on a regular desktop, to be compared with
times of the order of a week required by standard MD to exctract the
same amount of information. Moreover,
we have shown that this theoretical tool provides important new  insight
into the protein folding problem. In fact, it allows to define, characterize
and study the native and transition states and to determine the transition
time at different temperatures. We have also exhibited, within this framework,
a clear connection between the stability of the
folded conformation and its small conformational entropy. 

Applications of this formalism to the study of the conformational transitions of
small proteins with all-atom models and implicit solvent are in progress. 

\begin{acknowledgements}
Calculations were partly performed on the HPC facility "Wiglaf" at the Physics
Department of the University of Trento.
\end{acknowledgements}

\end{document}